\documentclass[twocolumn,showpacs,preprintnumbers,amsmath,amssymb, prl, reprint]{revtex4-1}

\usepackage{stmaryrd}
\usepackage{txfonts}
\usepackage{amssymb}
\usepackage{mathrsfs}
\usepackage{graphicx}
\usepackage{dcolumn}
\usepackage{bm}
\usepackage{epsfig}
\usepackage{color}                    
\usepackage{hyperref}                 

\begin{document}

\preprint{\href{http://dx.doi.org/10.1063/1.4809751}{S.-Z. Lin, C. Reichhardt and A. Saxena, Appl. Phys. Lett. {\bf 102}, 222405 (2013).}}

\title{Manipulation of skyrmions in nanodisks with a current pulse and skyrmion rectifier}

\author{Shi-Zeng Lin}
\affiliation{Theoretical Division, Los Alamos National Laboratory, Los Alamos, New Mexico 87545, USA}

\author{Charles Reichhardt}
\affiliation{Theoretical Division, Los Alamos National Laboratory, Los Alamos, New Mexico 87545, USA}

\author{Avadh Saxena}
\affiliation{Theoretical Division, Los Alamos National Laboratory, Los Alamos, New Mexico 87545, USA}

\begin{abstract}
A skyrmion in a nanosized disk of a chiral magnet can be used as a bit of information. To this end, it is desirable to control the creation and removal of a skyrmion only by currents without using external magnetic fields. Here we propose to create a skyrmion by applying a current pulse to a nanodisk. The skyrmion can be removed from the disk by applying a dc current. We show that the dynamics of the created skyrmion can lead to a rectification effect, in which a dc voltage is generated by the motion of skyrmion in the presence of an ac current.
\end{abstract}
\date{\today}
\maketitle

A new spin texture called skyrmion has been found in certain chiral magnets without inversion symmetry, such as MnSi and $\rm{Fe_{0.5}Co_{0.5}Si}$. \cite{Muhlbauer2009,Munzer10,Pfleiderer10,Yu2010a,Yu2011,Heinze2011,Seki2012} 
A skyrmion can be described in terms of spins which wrap a sphere. When a conduction electron passes through the skyrmion, its spin is fully polarized by the spin texture of the skyrmion, yielding a quantized Berry phase $4\pi$ that corresponds to a quantized magnetic flux $\Phi=hc/e$. This emergent electromagnetic field significantly affects the dynamics of the 
skyrmions and produces a Magnus force perpendicular to the skyrmion velocity. \cite{Iwasaki2013,szlin13skyrmion2} Application of a spin polarized current exerts a spin transfer torque on the skyrmion, causing it to move at an angle to the current direction determined by the dissipation. \cite{Zang11} The resulting voltage is perpendicular to the velocity; thus, a Hall (transverse) voltage is induced, which has been measured experimentally. \cite{Schulz2012}

Since skyrmions are stable topological excitations in chiral magnets, they have promising applications in spintronics. \cite{Fert2013} Due to the presence of the Magnus force \cite{Iwasaki2013,szlin13skyrmion2}, the threshold current to drive the skyrmion from the pinning sites is extremely small, $4-5$ orders of magnitude smaller than that for magnetic domain walls. \cite{Jonietz2010,Yu2012,Schulz2012} Thus, skyrmions have tremendous advantages for information storage applications, because the lower current implies less Joule heating. From a technological point of view, it is desirable to be able to manipulate, create, or destroy skyrmions by applying only a current without using an external magnetic field. This requires use of magnetic disk that is is typically the same size as the skyrmion, of the order of $100$ nm, such that only one skyrmion can be accommodated in the disk. The goal of producing a single magnetic vortex in a nanodisk was 
recently achieved \cite{Yamada2007,Kim2007}.  In most known chiral magnets, the skyrmion is not stable without an external magnetic field, and in the absence of a field a spiral structure is favored. Upon increasing the magnetic field, a skyrmion state becomes the ground state. \cite{Rossler2011} Two immediate questions regarding the dynamic creation of skyrmions arise: first, how to stabilize skyrmions without magnetic fields, and secondly, how to create a skyrmion using currents.

The answer to the first question is to exploit the magnetic anisotropy in thin films. \cite{Johnson1996} Due to the reduced dimensionality, an out-of-plane anisotropy develops that is inversely proportional to the film thickness. Such anisotropy to some extent acts as a perpendicular external field, so that skyrmions are easily stabilized in thin films. \cite{Yi09,Butenko2010,Yu2011,Wilson2012,Huang2012} For the second question, it was demonstrated in Ref. \onlinecite{Tchoe12} that skyrmions can be created by applying a circular current, which acts as an effective local magnetic field. However, implementation of a circular current may be difficult 
in many practical applications, especially for disk sizes that are below one micrometer. Here we introduce a mechanism of destabilizing the ferromagnetic (FM) state with a transport current, and then creating a metastable skyrmion.

We also explore the dynamics of the created skyrmion in a nanodisk. The Magnus force $\mathbf{F}_m\sim \hat{z}\times \mathbf{v}$ is always perpendicular to the velocity, so that the effective dynamics of the skyrmion is equivalent to a charged particle moving in a transverse magnetic field. Here $\hat{z}$ is a unit vector perpendicular to the disk. The trajectory of the skyrmion is thus  chiral, and is either clockwise or counterclockwise depending on the sign of the Magnus force. Under an ac current, we show that the trajectory of the skyrmion is an ellipse in the presence of a circularly symmetric confining potential. Since the voltage is perpendicular to the velocity, one thus can use skyrmions to extract a dc voltage by applying an ac current, indicating that it is possible to create
a \emph{skyrmion rectifier}.

\begin{figure}[t]
\psfig{figure=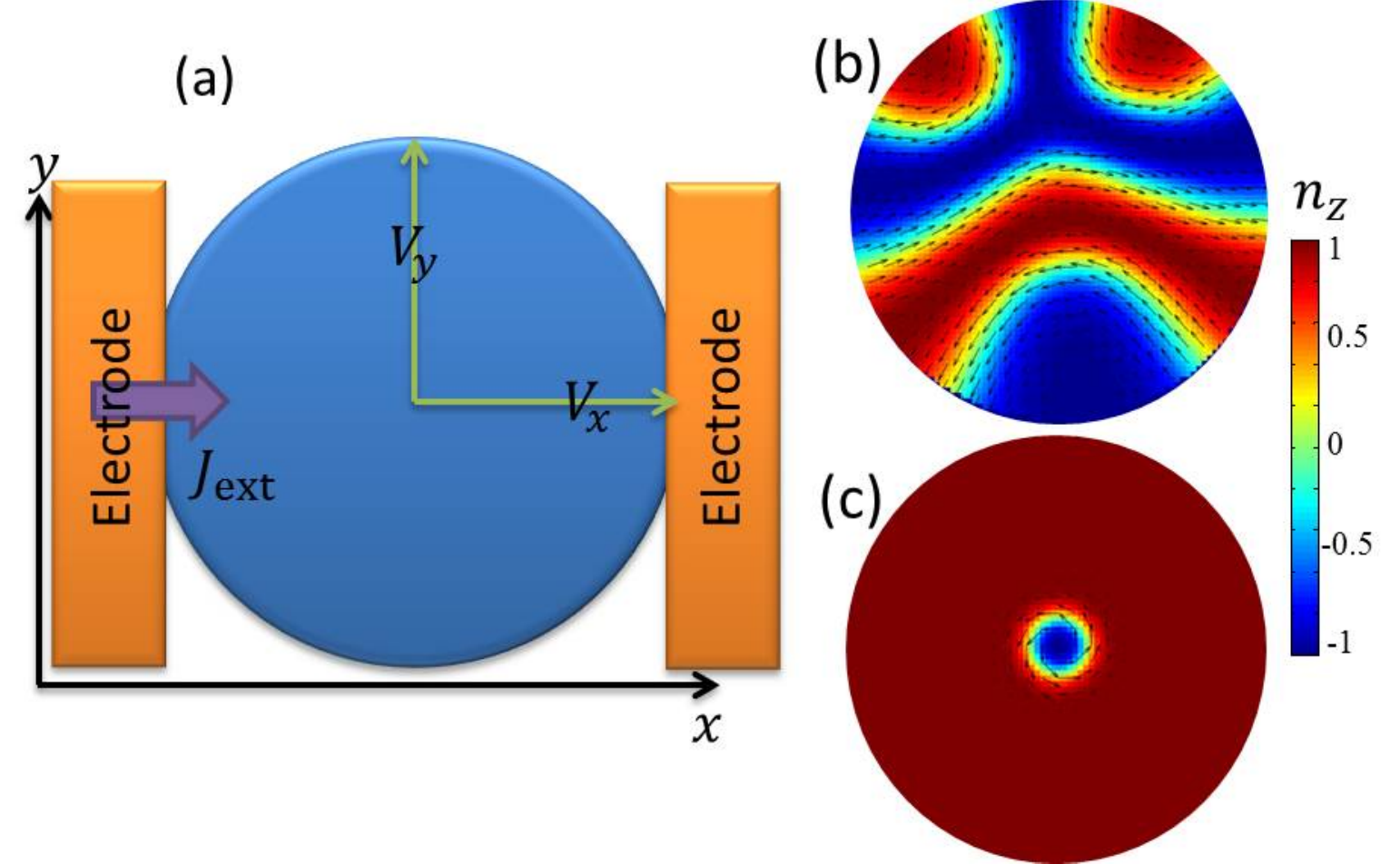,width=\columnwidth}
\caption{\label{f1}(color online) (a) Schematic view of a chiral magnet subjected to a current. To study the skyrmion rectification effect, we measure the voltage along the lines shown in the figure when an ac current is injected. (b) A twisted magnetic spiral at $J_A=0.4 J_{\text{ex}}/D$ and (c) a metastable skyrmion at $J_A=3.4 J_{\text{ex}}/D$. The vectors in the plots denote the $n_x$ and $n_y$ components, and color denotes the $n_z$ component. Here the radius is $R=8.0J_{\text{ex}}/D$.}
\end{figure}

We consider a thin circular disk of chiral magnet with typical size of the order of $100$ nm, comparable to the size of a single skyrmion. The system is described by the Hamiltonian \cite{Bogdanov89,Bogdanov94,Rosler2006,Han10}
\begin{equation}\label{eq1}
\mathcal{H}=d\int d\mathbf{r}^2 \left[\frac{J_{\rm{ex}}}{2}(\nabla \mathbf{n})^2+D\mathbf{n}\cdot\nabla\times \mathbf{n}-\frac{J_A}{2}n_z^2 \right],
\end{equation} 
where the first term is the exchange interaction, the second term is the Dzyaloshinskii-Moriya (DM) interaction due to the spin-orbit coupling \cite{Dzyaloshinsky1958,Moriya60,Moriya60b}, and the third term is the magnetic anisotropy with the easy axis perpendicular to the disk. For a thin film of thickness $d$, we have $J_A\propto 1/d$. \cite{Johnson1996} At low temperatures, the amplitude of the magnetization $\mathbf{M}$ is approximately fixed and is normalized in units of $\mathbf{n}=\mathbf{M}/M_s$ with $M_s$ the saturation magnetization. The dynamics of the spins are governed by the Landau-Lifshitz-Gilbert equation of motion \cite{Bazaliy98,Li04,Tatara2008}  
\begin{equation}\label{eq2}
{\partial _t}{\bf{n}} = \frac{\hbar\gamma}{2e}({{\bf{J}} }\cdot\nabla) {\bf{n}} - \gamma {\bf{n}} \times {{\bf{H}}_{\rm{eff}}} + \alpha {\partial _t}{\bf{n}} \times {\bf{n}},
\end{equation}
where the effective local field is $\mathbf{H}_{\rm{eff}}=-\delta\mathcal{H}/\delta \mathbf{n}=J_{\rm{ex}}{\nabla ^2}{\bf{n}} - 2D\nabla  \times {\bf{n}} +J_A n_z\hat{z}$. The first term is the spin transfer torque and the last term is the Gilbert damping with coefficient $\alpha$. Here $\gamma=a^3/(\hbar s)$ with $a$ the lattice constant and $s$ the total spin. The ground state can be found by numerical annealing by adding a Gaussian noise field to $\mathbf{H}_{\rm{eff}}$. For weak anisotropy, the ground state is a magnetic spiral, as shown in Fig. \ref{f1}(b). In the confined geometry, the direction of the spiral is twisted. Upon increasing the anisotropy, a FM state is stabilized. In this system without an external field, the out-of-plane anisotropy does not stabilize a skyrmion in the ground state. The skyrmion can, however, be a metastable state. We initially put a skyrmion at the center of the disk, and find in our simulation \cite{noteNumerics} that the skyrmion is stabilized for $2.5D^2/J_{\text{ex}}\lesssim J_A \lesssim 4.5D^2/J_{\text{ex}}$. For a large $J_A$, the size of the skyrmion shrinks because the whirl of spins would cost more energy in this case. The topological charge for the skyrmion is $Q=\int d\mathbf{r}^2q(\mathbf{r})$ with the skyrmion density $q = {\bf{n}}\cdot({\partial _x}{\bf{n}} \times {\partial _y}{\bf{n}})/(4\pi)$. \cite{SimonsQFT} The electric field induced by the motion of the skyrmion is $\mathbf{E}={\hbar }\mathbf{n}\cdot \left(\nabla \mathbf{n}\times\partial_t\mathbf{n} \right)/{2 e}$. \cite{Zang11}

We next study the creation of a skyrmion by a current pulse. The following conditions must be met. First, the anisotropy $J_A$ should be in the range $2.5D^2/J_{\text{ex}}\lesssim J_A \lesssim 4.5D^2/J_{\text{ex}}$ to support a metastable skyrmion excitation. Secondly, the amplitude of the current must be large enough to trigger the instability of FM state, 
as detailed below. Thirdly, the duration of the pulse should be large enough to allow the system to evolve into the skyrmion state. Typically this time is of the order of $10 J_{\rm{ex}}/(\gamma D^2)$. Finally, the damping constant should be moderately large. In simulations a skyrmion can be created easily with $\alpha>0.1$. For small damping, it is more difficult to stabilize the created skyrmion inside the disk because the skyrmion escapes from the disk during the time evolution.

\begin{figure*}[t]
\psfig{figure=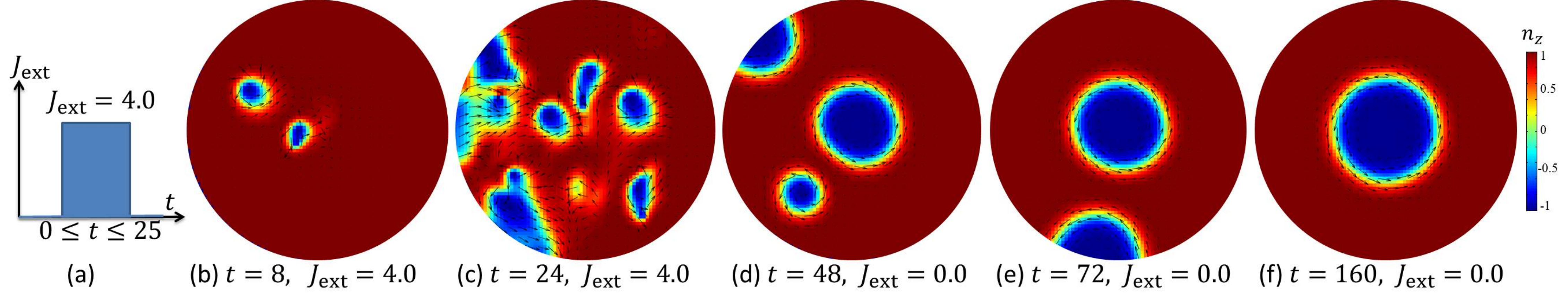,width=18cm}
\caption{\label{f2}(color online) 
(a) Profile of the current pulse used in simulations. (b)-(f) Time evolution of the spin texture in the disk. The vectors in the plots denote the $n_x$ and $n_y$ components, and color denotes the $n_z$ component. (b) and (c): After applying the current pulse, an instability in the FM state is triggered. (d): Three skyrmions are created after turning off the pulse, with the center of one skyrmion outside the disk. (e): The skyrmion at the edge in (c) leaves the disk and the other skyrmion is repelled towards the edges due to the inter-skyrmion repulsion. (f): Finally, one skyrmion is stabilized at the center of the disk. The radius of the disk is $R=8.0 J_{\text{ex}}/D$ and $J_A=3.0 D^2/J_{\text{ex}}$. The time in the plot is in units of $\tau=J_{\rm{ex}}/(\gamma D^2)$ and current is in units of $J_0=2D e/\hbar$. For MnSi, $\tau\approx 0.02$ ns and $J_0\approx 10^{12}\ \rm{A/m^2}$. To pin the created skyrmion, we introduce a defect at the center of the disk by assuming $J_{\rm{ex}}(\mathbf{r})=\bar{J}_{\rm{ex}}[1-0.8\exp(-r)]$. In simulations, an arbitrarily weak Gaussian noise field is added to $\mathbf{H}_{\rm{eff}}$ to trigger the instability.}
\end{figure*}

It is possible to create a skyrmion using current $\mathbf{J}$ because the FM state can be destabilized by a spin transfer torque caused by an applied  spin polarized current. \cite{szlin13skyrmion1} The spectrum of the spin wave excitation in the presence of a dc current $\mathbf{J}$ and the dissipation is
\begin{equation}\label{eq3}
\Omega =\mathbf{J}\cdot \mathbf{k}+ \frac{\gamma (1+i\alpha) }{\alpha ^2+1}\left(J_A+ J_{\rm{ex}} \mathbf{k}^2\right).
\end{equation}
The imaginary part of ${\rm{Im}}[\Omega]$ is due to the Gilbert damping. The spectrum becomes gapless when $J$ is larger than $J>J_i={2 \gamma  \sqrt{J_A J_{\text{ex}}}}/({\alpha ^2+1})$, which indicates an instability of the FM state. In the presence of weak perturbations, such as thermal/quantum fluctuations of spins, the perturbations increase exponentially. The external current couples to the emergent vector potential $\mathbf{A}$ associated with spin $\mathbf{n}$, where $\mathbf{A}\equiv {i c \hbar  b^{\dagger } \nabla b}/{e}$ with $b$ the spin coherent state \cite{SimonsQFT}. Thus the state with a non-zero vector potential $\mathbf{A}$ is favored after the ferromagnetic state has been made unstable. In the presence of the DM interaction, the state with a non-zero emergent vector potential is the skyrmion state. 

To demonstrate the dynamical creation of skyrmions by applying a current pulse, we perform numerical simulations of Eqs. \eqref{eq1} and \eqref{eq2} by adding an arbitrarily weak white noise magnetic field to $\mathbf{H}_{\mathrm{eff}}$. For numerical details, see Ref. \cite{noteNumerics}. During the time evolution when $J>J_i$ as shown in Fig. \ref{f2}, the $x$ or $y$ component of the spin increases, and a skyrmion is created in the final state. The created skyrmion is then driven by the current, moves towards the edge of the disk, and finally disappears. This processes repeats if a dc current is applied. In the presence of a current pulse, the created skyrmion will remain inside the disk if the current is turned off before the skyrmion has time to leave the disk. 

The nanodisk provides a geometric confinement that originates from the boundary condition, which prevents it from leaving the disk when no current is injected. This geometric confinement is effective only when the skyrmion is near the edge of the disk. In simulations, we can introduce a pinning center by modulating the electronic density at the center of the disk, since the exchange interaction is proportional to the electronic density. When $J_{\rm{ex}}(\mathbf{r})$ is depressed in the center of the disk, it provides a pinning potential for skyrmion. \cite{szlin13skyrmion2} In the case when the center of the skyrmion is already outside the disk, the skyrmion will be pushed away from the disk even though the current is turned off. In this situation, one may repeat the application of a current pulse with a slightly different duration to create a skyrmion. The removal of the skyrmion can be achieved by applying a dc current with amplitude smaller than $J_i$ but larger than the depinning current corresponding to the pinning potential. In this way, one can create and remove a skyrmion in a nanodisk in a controlled way using only a current. 

In the absence of an external magnetic field, the states obtained by reversing all the spins along the easy axis (the $z$ axis here) are degenerate. Therefore there are two types of skyrmions. In one type of skyrmion configuration, the spins in the skyrmion center are down and the spins away from the center are up (type I). The other degenerate skyrmion has the reverse spin direction, i.e., spins in the skyrmion center are up and the spins away from the center are down (type II). However, the chirality for both skyrmion configurations is the same, since it is determined by the sign of the DM interaction. In simulations, we found that if one starts from the FM state with spin up, one creates a skyrmion of type I, and vice versa.

Here we investigate the rectification effect, or the conversion of an ac current into a dc voltage, that originates from the unique dynamics of skyrmions in a nanodisk. To demonstrate the idea, we consider the particle-level description of a skyrmion in the presence of a pinning potential in dimensionless form \cite{szlin13skyrmion2}
\begin{equation}\label{eq4}
\eta\mathbf{v}=\hat{z}\times\mathbf{v}+\hat{z}\times\mathbf{J}+\mathbf{F}_d(\mathbf{r}_j-\mathbf{r}_i),
\end{equation}
where the first term on the right-hand side is the Magnus force, the second term is the Lorentz force, and the last term is the pinning force $\mathbf{F}_d=-\nabla U(\mathbf{r}_j-\mathbf{r}_i)$ with a pinning potential $U(\mathbf{r}_j-\mathbf{r}_i)$. Here the damping coefficient $\eta$ has the contributions from the Gilbert damping and the dissipation due to the conduction electrons.
The unique feature of the dynamic equation is the presence of the Magnus force, which causes the skyrmion motion to be clockwise in the $x-y$ plane. The trajectory of a skyrmion moving in a strong pinning potential and subject to an ac current $J_x=J_{\rm{x0}} \sin \omega t$ forms a limit cycle, according to the Poincar\'{e}-Bendixson theorem, \cite{StrogatzBook} because (1) there is no fixed point; and (2) the skyrmion trajectory is bounded by the pinning potential. For a parabolic pinning potential $U=U_p r^2/2$, the trajectory of the skyrmion can be found analytically and is an ellipse: 
\begin{equation}
x(t)={\rm{Re}}\left[\frac{i \omega  J_{\text{x0}}}{\eta ^2 \omega ^2-2 i \eta  \omega  U_p-U_p^2+\omega ^2} \exp(i\omega t)\right],
\end{equation}
\begin{equation}
y(t)={\rm{Re}}\left[-\frac{J_{\text{x0}} U_p+i \eta  \omega  J_{\text{x0}}}{\eta ^2 \omega ^2-2 i \eta  \omega  U_p-U_p^2+\omega ^2} \exp(i\omega t)\right].
\end{equation}
When $U_p\approx \omega$ and $\eta\ll 1$, the trajectory is nearly circular, and when $\eta\gg 1$, it is nearly linear. If one measures the voltage along the radial direction $V_r$, which is proportional to the skyrmion velocity in the azimuthal direction $V_r\propto v_{\phi}$, one can extract a dc voltage even though only an ac current is applied, i.e. 
a skyrmion rectifier appears. 

\begin{figure}[b]
\psfig{figure=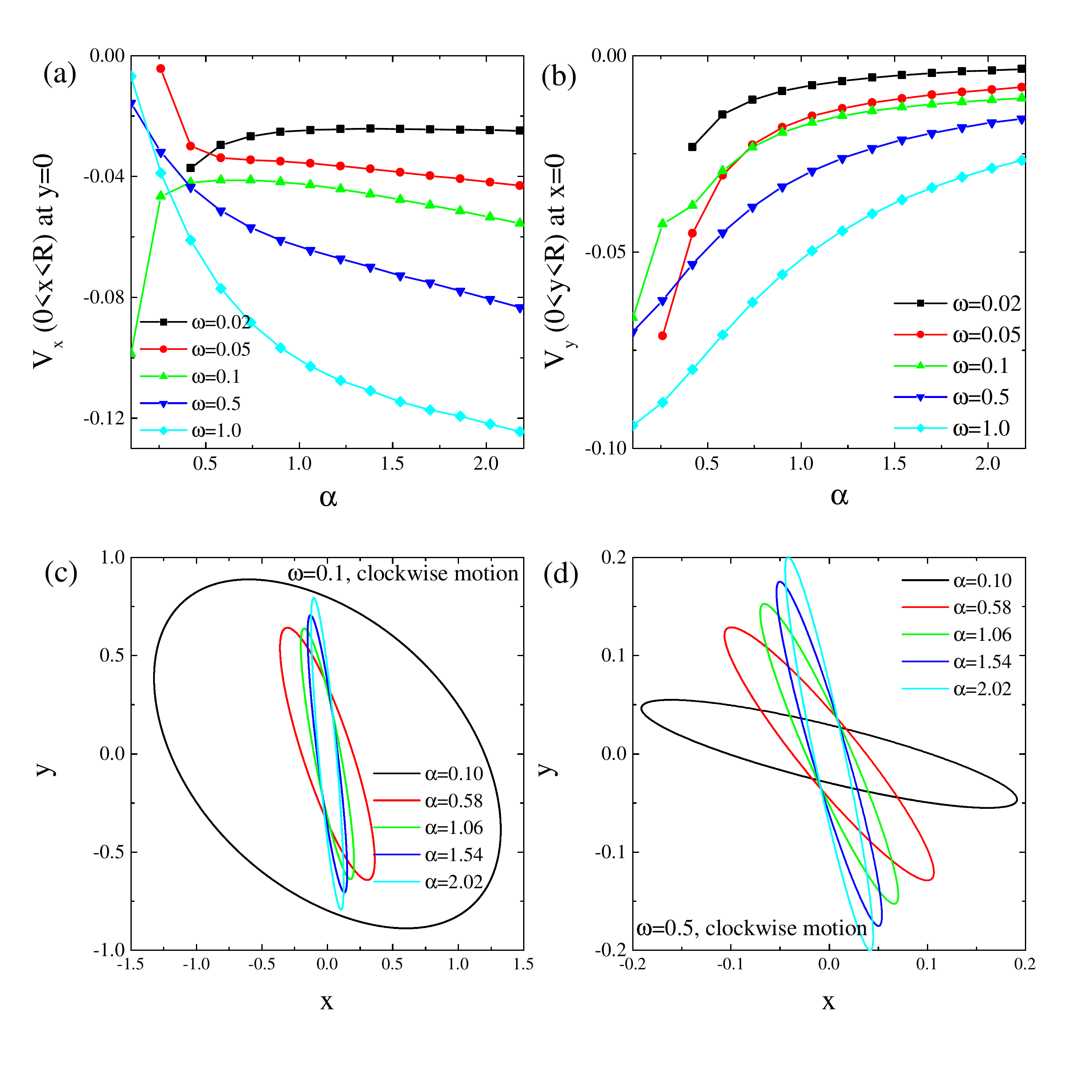,width=\columnwidth}
\caption{\label{f3}(color online) Dependence on $\alpha$ of (a) the radial voltage $V_x$ measured for $0<x<R$ at $y=0$ and (b) the radial voltage $V_y$ measured for $0<y<R$ at $x=0$ at different driving frequencies $\omega$ of the ac current. (c) and (d) The corresponding trajectories of skyrmions at different $\alpha$ and $\omega$. An ac current along the $x$ direction is injected 
with the form $J_x=J_{\rm{x0}}\sin(\omega t)$, where the amplitude $J_{x0}=0.2D e/\hbar$. A defect is introduced at the center of the disk by assuming $J_{\rm{ex}}(\mathbf{r})=\bar{J}_{\rm{ex}}[1-0.8\exp(-r)]$.}
\end{figure}

We consider a nanodisk described by the Hamiltonian in Eq. \eqref{eq1}, which supports a skyrmion as a metastable state. We explicitly introduce a defect at the center of disk to pin the skyrmion by assuming $J_{\rm{ex}}(\mathbf{r})=\bar{J}_{\rm{ex}}[1-0.8\exp(-r)]$. Initially we put a skyrmion in the center of the disk and add an ac current with amplitude smaller than the depinning current, such that the skyrmion is not driven outside the disk. We calculate the time averaged voltage between ($R$, 0) and (0, 0), as well as the voltage between (0, $R$) and (0, 0), as shown in Fig. \ref{f1}, by solving Eqs. \eqref{eq1} and \eqref{eq2} numerically \cite{noteNumerics}. We also calculate the time dependence of the center of mass defined as
\begin{equation}
\mathbf{r}_c=\frac{1}{Q}\int d \mathbf{r}^2 q(\mathbf{r})\mathbf{r}.
\end{equation}
The trajectories at several different values of $\alpha$ are shown in Fig. \ref{f3}. As $\alpha$ increases the dissipative force becomes dominant over the Mangus force, causing the trajectory to become elliptical. In the $\alpha \rightarrow \infty$ limit, the trajectory becomes a straight line in the radial direction and thus the time averaged dc voltage vanishes.  The dependence of the voltage on $\alpha$ is shown in Fig. \ref{f3}. There is a clear correspondence between the voltage and the skyrmion trajectory. For instance, as $\alpha$ increases, the larger principal axis of the ellipse approaches the $y$ axis. Thus the amplitude of $V_x (0<x<R)$ increases while $V_y (0<y<R)$ decreases in most cases. Thus the voltage in Figs. \ref{f3} (a) and (b) can be inferred from the trajectories in Figs. \ref{f3} (c) and (d).  Particularly for $\omega=0.1$, the dependence of $V_x$ on $\alpha$ is non-monotonic. All these behaviors are captured by the particle model in Eq. \eqref{eq4}.

We use typical parameters for MnSi to estimate the optimal parameters for the creation of a skyrmion. For MnSi, $J_{\rm{ex}}\approx 3\ {\rm{meV}}/a$ and $D\approx 0.3\ {\rm{meV}}/a^2$ with the lattice constant $a\approx 2.9\ \AA$.  \cite{Zang11} For a disk size of about $100$ nm, the strength of the current pulse should be larger than $J>J_i\sim 10^{12}\ \rm{A/m^2}$, while the duration should be longer than $0.2$ ns. To reach the range of the uniaxial anisotropy $2.5D^2/J_{\text{ex}}\lesssim J_A \lesssim 4.5D^2/J_{\text{ex}}$ required for stabilizing a skyrmion, the thickness of the disk should be around $d\approx 5$ nm. \cite{Johnson1996} One challenge here is to enhance the damping coefficient $\alpha>0.1$ for a skyrmion, since the bare Gilbert damping is usually very weak, $\alpha_G<0.1$. Nevertheless, there is an additional damping mechanism due to the electric field induced by the motion of a skyrmion. \cite{Zang11} The dressed damping thus can be larger than $\alpha>0.1$, which is optimal for the creation of skyrmions.  

Finally, we estimate the conversion efficiency of the skyrmion rectifier, which is defined as the ratio of the induced dc voltage $V_{\text{dc}}$ to the amplitude of the ac current $P\equiv V_{\text{dc}}/J_{\rm{x0}}$. According to the results in Fig. \ref{f3} for a single skyrmion, we obtain $P\approx 10^{-17}\ \rm{V\cdot m^2/A}$. The rectification effect is weak; however, it is experimentally observable. For an applied current $J_{\rm{x0}}\approx 10^{12}\ \rm{A/m^2}$, we have $V_{\text{dc}}\approx 10\ \rm{\mu V}$. Note that $V_{\text{dc}}$ can be enhanced at a given $J_{\rm{x0}}$ by using many skyrmions in a large disk, since the voltage is proportional to the number of skyrmions. In a large thin film without pinning centers, $V_{\rm{dc}}$ is just the Hall voltage associated with the motion of skyrmions, which has been measured experimentally. \cite{Schulz2012}

The applied current generates a magnetic field and we estimate its strength. For a current density $10^{12}\ \rm{A/m^2}$, disk radius $100$ nm and disk thickness $5$ nm, the induced field is about $25$ G at the surface of the disk. The generated field is much weaker than that used in experiments to stabilize the skyrmion phase. Thus we expect the self-induced magnetic field has a negligible effect on the dynamical creation and removal of skyrmions, and also the skyrmion dynamics discussed here. 

To summarize, we have demonstrated a method to create and annihilate skyrmions in chiral magnetic nanodisks without an applied magnetic field by utilizing an applied current to create an instability in the ferromagnetic state. The ability to stabilize skyrmions in nanodisks would be a significant step forward for possible applications of skyrmions. Additionally, we show that once a skyrmion is stabilized in a disk, the unique skyrmion dynamics produce complex skyrmion orbits under the application of an ac drive, leading to the production of a dc voltage similar to a rectification effect.

\vspace{2mm}
\acknowledgments
We thank Cristian D. Batista, Boris A. Maiorov, and Yasuyuki Kato for useful discussions and Cynthia J.O. Reichhardt for a critical reading of the manuscript. This publication was made possible by funding from the Los Alamos Laboratory Directed Research and Development Program, project number 20110138ER. This work was carried out under the auspices of the NNSA of the U.S. DoE at LANL under Contract No. DE-AC52-06NA25396.


\begin{thebibliography}{38}%
\makeatletter
\providecommand \@ifxundefined [1]{%
 \@ifx{#1\undefined}
}%
\providecommand \@ifnum [1]{%
 \ifnum #1\expandafter \@firstoftwo
 \else \expandafter \@secondoftwo
 \fi
}%
\providecommand \@ifx [1]{%
 \ifx #1\expandafter \@firstoftwo
 \else \expandafter \@secondoftwo
 \fi
}%
\providecommand \natexlab [1]{#1}%
\providecommand \enquote  [1]{``#1''}%
\providecommand \bibnamefont  [1]{#1}%
\providecommand \bibfnamefont [1]{#1}%
\providecommand \citenamefont [1]{#1}%
\providecommand \href@noop [0]{\@secondoftwo}%
\providecommand \href [0]{\begingroup \@sanitize@url \@href}%
\providecommand \@href[1]{\@@startlink{#1}\@@href}%
\providecommand \@@href[1]{\endgroup#1\@@endlink}%
\providecommand \@sanitize@url [0]{\catcode `\\12\catcode `\$12\catcode
  `\&12\catcode `\#12\catcode `\^12\catcode `\_12\catcode `\%12\relax}%
\providecommand \@@startlink[1]{}%
\providecommand \@@endlink[0]{}%
\providecommand \url  [0]{\begingroup\@sanitize@url \@url }%
\providecommand \@url [1]{\endgroup\@href {#1}{\urlprefix }}%
\providecommand \urlprefix  [0]{URL }%
\providecommand \Eprint [0]{\href }%
\providecommand \doibase [0]{http://dx.doi.org/}%
\providecommand \selectlanguage [0]{\@gobble}%
\providecommand \bibinfo  [0]{\@secondoftwo}%
\providecommand \bibfield  [0]{\@secondoftwo}%
\providecommand \translation [1]{[#1]}%
\providecommand \BibitemOpen [0]{}%
\providecommand \bibitemStop [0]{}%
\providecommand \bibitemNoStop [0]{.\EOS\space}%
\providecommand \EOS [0]{\spacefactor3000\relax}%
\providecommand \BibitemShut  [1]{\csname bibitem#1\endcsname}%
\let\auto@bib@innerbib\@empty
\bibitem [{\citenamefont {M\"{u}hlbauer}\ \emph {et~al.}(2009)\citenamefont
  {M\"{u}hlbauer}, \citenamefont {Binz}, \citenamefont {Jonietz}, \citenamefont
  {Pfleiderer}, \citenamefont {Rosch}, \citenamefont {Neubauer}, \citenamefont
  {Georgii},\ and\ \citenamefont {B\"{o}ni}}]{Muhlbauer2009}%
  \BibitemOpen
  \bibfield  {author} {\bibinfo {author} {\bibfnamefont {S.}~\bibnamefont
  {M\"{u}hlbauer}}, \bibinfo {author} {\bibfnamefont {B.}~\bibnamefont {Binz}},
  \bibinfo {author} {\bibfnamefont {F.}~\bibnamefont {Jonietz}}, \bibinfo
  {author} {\bibfnamefont {C.}~\bibnamefont {Pfleiderer}}, \bibinfo {author}
  {\bibfnamefont {A.}~\bibnamefont {Rosch}}, \bibinfo {author} {\bibfnamefont
  {A.}~\bibnamefont {Neubauer}}, \bibinfo {author} {\bibfnamefont
  {R.}~\bibnamefont {Georgii}}, \ and\ \bibinfo {author} {\bibfnamefont
  {P.}~\bibnamefont {B\"{o}ni}},\ }\href {\doibase 10.1126/science.1166767}
  {\bibfield  {journal} {\bibinfo  {journal} {Science}\ }\textbf {\bibinfo
  {volume} {323}},\ \bibinfo {pages} {915} (\bibinfo {year}
  {2009})}\BibitemShut {NoStop}%
\bibitem [{\citenamefont {M\"unzer}\ \emph {et~al.}(2010)\citenamefont
  {M\"unzer}, \citenamefont {Neubauer}, \citenamefont {Adams}, \citenamefont
  {M\"uhlbauer}, \citenamefont {Franz}, \citenamefont {Jonietz}, \citenamefont
  {Georgii}, \citenamefont {B\"oni}, \citenamefont {Pedersen}, \citenamefont
  {Schmidt}, \citenamefont {Rosch},\ and\ \citenamefont
  {Pfleiderer}}]{Munzer10}%
  \BibitemOpen
  \bibfield  {author} {\bibinfo {author} {\bibfnamefont {W.}~\bibnamefont
  {M\"unzer}}, \bibinfo {author} {\bibfnamefont {A.}~\bibnamefont {Neubauer}},
  \bibinfo {author} {\bibfnamefont {T.}~\bibnamefont {Adams}}, \bibinfo
  {author} {\bibfnamefont {S.}~\bibnamefont {M\"uhlbauer}}, \bibinfo {author}
  {\bibfnamefont {C.}~\bibnamefont {Franz}}, \bibinfo {author} {\bibfnamefont
  {F.}~\bibnamefont {Jonietz}}, \bibinfo {author} {\bibfnamefont
  {R.}~\bibnamefont {Georgii}}, \bibinfo {author} {\bibfnamefont
  {P.}~\bibnamefont {B\"oni}}, \bibinfo {author} {\bibfnamefont
  {B.}~\bibnamefont {Pedersen}}, \bibinfo {author} {\bibfnamefont
  {M.}~\bibnamefont {Schmidt}}, \bibinfo {author} {\bibfnamefont
  {A.}~\bibnamefont {Rosch}}, \ and\ \bibinfo {author} {\bibfnamefont
  {C.}~\bibnamefont {Pfleiderer}},\ }\href {\doibase
  10.1103/PhysRevB.81.041203} {\bibfield  {journal} {\bibinfo  {journal} {Phys.
  Rev. B}\ }\textbf {\bibinfo {volume} {81}},\ \bibinfo {pages} {041203}
  (\bibinfo {year} {2010})}\BibitemShut {NoStop}%
\bibitem [{\citenamefont {Pfleiderer}\ \emph {et~al.}(2010)\citenamefont
  {Pfleiderer}, \citenamefont {Adams}, \citenamefont {Bauer}, \citenamefont
  {Biberacher}, \citenamefont {Binz}, \citenamefont {Birkelbach}, \citenamefont
  {B\"{o}ni}, \citenamefont {Franz}, \citenamefont {Georgii}, \citenamefont
  {Janoschek}, \citenamefont {Jonietz}, \citenamefont {Keller}, \citenamefont
  {Ritz}, \citenamefont {M\"uhlbauer}, \citenamefont {M\"unzer}, \citenamefont
  {Neubauer}, \citenamefont {Pedersen},\ and\ \citenamefont
  {Rosch}}]{Pfleiderer10}%
  \BibitemOpen
  \bibfield  {author} {\bibinfo {author} {\bibfnamefont {C.}~\bibnamefont
  {Pfleiderer}}, \bibinfo {author} {\bibfnamefont {T.}~\bibnamefont {Adams}},
  \bibinfo {author} {\bibfnamefont {A.}~\bibnamefont {Bauer}}, \bibinfo
  {author} {\bibfnamefont {W.}~\bibnamefont {Biberacher}}, \bibinfo {author}
  {\bibfnamefont {B.}~\bibnamefont {Binz}}, \bibinfo {author} {\bibfnamefont
  {F.}~\bibnamefont {Birkelbach}}, \bibinfo {author} {\bibfnamefont
  {P.}~\bibnamefont {B\"{o}ni}}, \bibinfo {author} {\bibfnamefont
  {C.}~\bibnamefont {Franz}}, \bibinfo {author} {\bibfnamefont
  {R.}~\bibnamefont {Georgii}}, \bibinfo {author} {\bibfnamefont
  {M.}~\bibnamefont {Janoschek}}, \bibinfo {author} {\bibfnamefont
  {F.}~\bibnamefont {Jonietz}}, \bibinfo {author} {\bibfnamefont
  {T.}~\bibnamefont {Keller}}, \bibinfo {author} {\bibfnamefont
  {R.}~\bibnamefont {Ritz}}, \bibinfo {author} {\bibfnamefont {S.}~\bibnamefont
  {M\"uhlbauer}}, \bibinfo {author} {\bibfnamefont {W.}~\bibnamefont
  {M\"unzer}}, \bibinfo {author} {\bibfnamefont {A.}~\bibnamefont {Neubauer}},
  \bibinfo {author} {\bibfnamefont {B.}~\bibnamefont {Pedersen}}, \ and\
  \bibinfo {author} {\bibfnamefont {A.}~\bibnamefont {Rosch}},\ }\href
  {http://stacks.iop.org/0953-8984/22/i=16/a=164207} {\bibfield  {journal}
  {\bibinfo  {journal} {J. Phys.: Condens. Matter}\ }\textbf {\bibinfo {volume}
  {22}},\ \bibinfo {pages} {164207} (\bibinfo {year} {2010})}\BibitemShut
  {NoStop}%
\bibitem [{\citenamefont {Yu}\ \emph {et~al.}(2010)\citenamefont {Yu},
  \citenamefont {Onose}, \citenamefont {Kanazawa}, \citenamefont {Park},
  \citenamefont {Han}, \citenamefont {Matsui}, \citenamefont {Nagaosa},\ and\
  \citenamefont {Tokura}}]{Yu2010a}%
  \BibitemOpen
  \bibfield  {author} {\bibinfo {author} {\bibfnamefont {X.~Z.}\ \bibnamefont
  {Yu}}, \bibinfo {author} {\bibfnamefont {Y.}~\bibnamefont {Onose}}, \bibinfo
  {author} {\bibfnamefont {N.}~\bibnamefont {Kanazawa}}, \bibinfo {author}
  {\bibfnamefont {J.~H.}\ \bibnamefont {Park}}, \bibinfo {author}
  {\bibfnamefont {J.~H.}\ \bibnamefont {Han}}, \bibinfo {author} {\bibfnamefont
  {Y.}~\bibnamefont {Matsui}}, \bibinfo {author} {\bibfnamefont
  {N.}~\bibnamefont {Nagaosa}}, \ and\ \bibinfo {author} {\bibfnamefont
  {Y.}~\bibnamefont {Tokura}},\ }\href {\doibase 10.1038/nature09124}
  {\bibfield  {journal} {\bibinfo  {journal} {Nature}\ }\textbf {\bibinfo
  {volume} {465}},\ \bibinfo {pages} {901} (\bibinfo {year}
  {2010})}\BibitemShut {NoStop}%
\bibitem [{\citenamefont {Yu}\ \emph {et~al.}(2011)\citenamefont {Yu},
  \citenamefont {Kanazawa}, \citenamefont {Onose}, \citenamefont {Kimoto},
  \citenamefont {Zhang}, \citenamefont {Ishiwata}, \citenamefont {Matsui},\
  and\ \citenamefont {Tokura}}]{Yu2011}%
  \BibitemOpen
  \bibfield  {author} {\bibinfo {author} {\bibfnamefont {X.~Z.}\ \bibnamefont
  {Yu}}, \bibinfo {author} {\bibfnamefont {N.}~\bibnamefont {Kanazawa}},
  \bibinfo {author} {\bibfnamefont {Y.}~\bibnamefont {Onose}}, \bibinfo
  {author} {\bibfnamefont {K.}~\bibnamefont {Kimoto}}, \bibinfo {author}
  {\bibfnamefont {W.~Z.}\ \bibnamefont {Zhang}}, \bibinfo {author}
  {\bibfnamefont {S.}~\bibnamefont {Ishiwata}}, \bibinfo {author}
  {\bibfnamefont {Y.}~\bibnamefont {Matsui}}, \ and\ \bibinfo {author}
  {\bibfnamefont {Y.}~\bibnamefont {Tokura}},\ }\href {\doibase
  10.1038/nmat2916} {\bibfield  {journal} {\bibinfo  {journal} {Nature
  Materials}\ }\textbf {\bibinfo {volume} {10}},\ \bibinfo {pages} {106}
  (\bibinfo {year} {2011})}\BibitemShut {NoStop}%
\bibitem [{\citenamefont {Heinze}\ \emph {et~al.}(2011)\citenamefont {Heinze},
  \citenamefont {Bergmann}, \citenamefont {Menzel}, \citenamefont {Brede},
  \citenamefont {Kubetzka}, \citenamefont {Wiesendanger}, \citenamefont
  {Bihlmayer},\ and\ \citenamefont {Blügel}}]{Heinze2011}%
  \BibitemOpen
  \bibfield  {author} {\bibinfo {author} {\bibfnamefont {S.}~\bibnamefont
  {Heinze}}, \bibinfo {author} {\bibfnamefont {K.~v.}\ \bibnamefont
  {Bergmann}}, \bibinfo {author} {\bibfnamefont {M.}~\bibnamefont {Menzel}},
  \bibinfo {author} {\bibfnamefont {J.}~\bibnamefont {Brede}}, \bibinfo
  {author} {\bibfnamefont {A.}~\bibnamefont {Kubetzka}}, \bibinfo {author}
  {\bibfnamefont {R.}~\bibnamefont {Wiesendanger}}, \bibinfo {author}
  {\bibfnamefont {G.}~\bibnamefont {Bihlmayer}}, \ and\ \bibinfo {author}
  {\bibfnamefont {S.}~\bibnamefont {Blügel}},\ }\href {\doibase
  10.1038/nphys2045} {\bibfield  {journal} {\bibinfo  {journal} {Nature
  Physics}\ }\textbf {\bibinfo {volume} {7}},\ \bibinfo {pages} {713} (\bibinfo
  {year} {2011})}\BibitemShut {NoStop}%
\bibitem [{\citenamefont {Seki}\ \emph {et~al.}(2012)\citenamefont {Seki},
  \citenamefont {Yu}, \citenamefont {Ishiwata},\ and\ \citenamefont
  {Tokura}}]{Seki2012}%
  \BibitemOpen
  \bibfield  {author} {\bibinfo {author} {\bibfnamefont {S.}~\bibnamefont
  {Seki}}, \bibinfo {author} {\bibfnamefont {X.~Z.}\ \bibnamefont {Yu}},
  \bibinfo {author} {\bibfnamefont {S.}~\bibnamefont {Ishiwata}}, \ and\
  \bibinfo {author} {\bibfnamefont {Y.}~\bibnamefont {Tokura}},\ }\href
  {\doibase 10.1126/science.1214143} {\bibfield  {journal} {\bibinfo  {journal}
  {Science}\ }\textbf {\bibinfo {volume} {336}},\ \bibinfo {pages} {198}
  (\bibinfo {year} {2012})}\BibitemShut {NoStop}%
\bibitem [{\citenamefont {Iwasaki}\ \emph {et~al.}(2013)\citenamefont
  {Iwasaki}, \citenamefont {Mochizuki},\ and\ \citenamefont
  {Nagaosa}}]{Iwasaki2013}%
  \BibitemOpen
  \bibfield  {author} {\bibinfo {author} {\bibfnamefont {J.}~\bibnamefont
  {Iwasaki}}, \bibinfo {author} {\bibfnamefont {M.}~\bibnamefont {Mochizuki}},
  \ and\ \bibinfo {author} {\bibfnamefont {N.}~\bibnamefont {Nagaosa}},\ }\href
  {\doibase 10.1038/ncomms2442} {\bibfield  {journal} {\bibinfo  {journal}
  {Nature Communications}\ }\textbf {\bibinfo {volume} {4}},\ \bibinfo {pages}
  {1463} (\bibinfo {year} {2013})}\BibitemShut {NoStop}%
\bibitem [{\citenamefont {Lin}\ \emph {et~al.}(2013{\natexlab{a}})\citenamefont
  {Lin}, \citenamefont {Reichhardt}, \citenamefont {Batista},\ and\
  \citenamefont {Saxena}}]{szlin13skyrmion2}%
  \BibitemOpen
  \bibfield  {author} {\bibinfo {author} {\bibfnamefont {S.~Z.}\ \bibnamefont
  {Lin}}, \bibinfo {author} {\bibfnamefont {C.}~\bibnamefont {Reichhardt}},
  \bibinfo {author} {\bibfnamefont {C.~D.}\ \bibnamefont {Batista}}, \ and\
  \bibinfo {author} {\bibfnamefont {A.}~\bibnamefont {Saxena}},\ }\href@noop {}
  {\bibfield  {journal} {\bibinfo  {journal} {arXiv:1302.6205}\ } (\bibinfo
  {year} {2013}{\natexlab{a}})}\BibitemShut {NoStop}%
\bibitem [{\citenamefont {Zang}\ \emph {et~al.}(2011)\citenamefont {Zang},
  \citenamefont {Mostovoy}, \citenamefont {Han},\ and\ \citenamefont
  {Nagaosa}}]{Zang11}%
  \BibitemOpen
  \bibfield  {author} {\bibinfo {author} {\bibfnamefont {J.}~\bibnamefont
  {Zang}}, \bibinfo {author} {\bibfnamefont {M.}~\bibnamefont {Mostovoy}},
  \bibinfo {author} {\bibfnamefont {J.~H.}\ \bibnamefont {Han}}, \ and\
  \bibinfo {author} {\bibfnamefont {N.}~\bibnamefont {Nagaosa}},\ }\href
  {\doibase 10.1103/PhysRevLett.107.136804} {\bibfield  {journal} {\bibinfo
  {journal} {Phys. Rev. Lett.}\ }\textbf {\bibinfo {volume} {107}},\ \bibinfo
  {pages} {136804} (\bibinfo {year} {2011})}\BibitemShut {NoStop}%
\bibitem [{\citenamefont {Schulz}\ \emph {et~al.}(2012)\citenamefont {Schulz},
  \citenamefont {Ritz}, \citenamefont {Bauer}, \citenamefont {Halder},
  \citenamefont {Wagner}, \citenamefont {Franz}, \citenamefont {Pfleiderer},
  \citenamefont {Everschor}, \citenamefont {Garst},\ and\ \citenamefont
  {Rosch}}]{Schulz2012}%
  \BibitemOpen
  \bibfield  {author} {\bibinfo {author} {\bibfnamefont {T.}~\bibnamefont
  {Schulz}}, \bibinfo {author} {\bibfnamefont {R.}~\bibnamefont {Ritz}},
  \bibinfo {author} {\bibfnamefont {A.}~\bibnamefont {Bauer}}, \bibinfo
  {author} {\bibfnamefont {M.}~\bibnamefont {Halder}}, \bibinfo {author}
  {\bibfnamefont {M.}~\bibnamefont {Wagner}}, \bibinfo {author} {\bibfnamefont
  {C.}~\bibnamefont {Franz}}, \bibinfo {author} {\bibfnamefont
  {C.}~\bibnamefont {Pfleiderer}}, \bibinfo {author} {\bibfnamefont
  {K.}~\bibnamefont {Everschor}}, \bibinfo {author} {\bibfnamefont
  {M.}~\bibnamefont {Garst}}, \ and\ \bibinfo {author} {\bibfnamefont
  {A.}~\bibnamefont {Rosch}},\ }\href {\doibase 10.1038/nphys2231} {\bibfield
  {journal} {\bibinfo  {journal} {Nature Physics}\ }\textbf {\bibinfo {volume}
  {8}},\ \bibinfo {pages} {301} (\bibinfo {year} {2012})}\BibitemShut {NoStop}%
\bibitem [{\citenamefont {Fert}\ \emph {et~al.}(2013)\citenamefont {Fert},
  \citenamefont {Cros},\ and\ \citenamefont {Sampaio}}]{Fert2013}%
  \BibitemOpen
  \bibfield  {author} {\bibinfo {author} {\bibfnamefont {A.}~\bibnamefont
  {Fert}}, \bibinfo {author} {\bibfnamefont {V.}~\bibnamefont {Cros}}, \ and\
  \bibinfo {author} {\bibfnamefont {J.}~\bibnamefont {Sampaio}},\ }\href
  {\doibase 10.1038/nnano.2013.29} {\bibfield  {journal} {\bibinfo  {journal}
  {Nature Nanotechnology}\ }\textbf {\bibinfo {volume} {8}},\ \bibinfo {pages}
  {152} (\bibinfo {year} {2013})}\BibitemShut {NoStop}%
\bibitem [{\citenamefont {Jonietz}\ \emph {et~al.}(2010)\citenamefont
  {Jonietz}, \citenamefont {M\"uhlbauer}, \citenamefont {Pfleiderer},
  \citenamefont {Neubauer}, \citenamefont {M\"unzer}, \citenamefont {Bauer},
  \citenamefont {Adams}, \citenamefont {Georgii}, \citenamefont {B\"oni},
  \citenamefont {Duine}, \citenamefont {Everschor}, \citenamefont {Garst},\
  and\ \citenamefont {Rosch}}]{Jonietz2010}%
  \BibitemOpen
  \bibfield  {author} {\bibinfo {author} {\bibfnamefont {F.}~\bibnamefont
  {Jonietz}}, \bibinfo {author} {\bibfnamefont {S.}~\bibnamefont
  {M\"uhlbauer}}, \bibinfo {author} {\bibfnamefont {C.}~\bibnamefont
  {Pfleiderer}}, \bibinfo {author} {\bibfnamefont {A.}~\bibnamefont
  {Neubauer}}, \bibinfo {author} {\bibfnamefont {W.}~\bibnamefont {M\"unzer}},
  \bibinfo {author} {\bibfnamefont {A.}~\bibnamefont {Bauer}}, \bibinfo
  {author} {\bibfnamefont {T.}~\bibnamefont {Adams}}, \bibinfo {author}
  {\bibfnamefont {R.}~\bibnamefont {Georgii}}, \bibinfo {author} {\bibfnamefont
  {P.}~\bibnamefont {B\"oni}}, \bibinfo {author} {\bibfnamefont {R.~A.}\
  \bibnamefont {Duine}}, \bibinfo {author} {\bibfnamefont {K.}~\bibnamefont
  {Everschor}}, \bibinfo {author} {\bibfnamefont {M.}~\bibnamefont {Garst}}, \
  and\ \bibinfo {author} {\bibfnamefont {A.}~\bibnamefont {Rosch}},\ }\href
  {\doibase 10.1126/science.1195709} {\bibfield  {journal} {\bibinfo  {journal}
  {Science}\ }\textbf {\bibinfo {volume} {330}},\ \bibinfo {pages} {1648}
  (\bibinfo {year} {2010})}\BibitemShut {NoStop}%
\bibitem [{\citenamefont {Yu}\ \emph {et~al.}(2012)\citenamefont {Yu},
  \citenamefont {Kanazawa}, \citenamefont {Zhang}, \citenamefont {Nagai},
  \citenamefont {Hara}, \citenamefont {Kimoto}, \citenamefont {Matsui},
  \citenamefont {Onose},\ and\ \citenamefont {Tokura}}]{Yu2012}%
  \BibitemOpen
  \bibfield  {author} {\bibinfo {author} {\bibfnamefont {X.~Z.}\ \bibnamefont
  {Yu}}, \bibinfo {author} {\bibfnamefont {N.}~\bibnamefont {Kanazawa}},
  \bibinfo {author} {\bibfnamefont {W.~Z.}\ \bibnamefont {Zhang}}, \bibinfo
  {author} {\bibfnamefont {T.}~\bibnamefont {Nagai}}, \bibinfo {author}
  {\bibfnamefont {T.}~\bibnamefont {Hara}}, \bibinfo {author} {\bibfnamefont
  {K.}~\bibnamefont {Kimoto}}, \bibinfo {author} {\bibfnamefont
  {Y.}~\bibnamefont {Matsui}}, \bibinfo {author} {\bibfnamefont
  {Y.}~\bibnamefont {Onose}}, \ and\ \bibinfo {author} {\bibfnamefont
  {Y.}~\bibnamefont {Tokura}},\ }\href {\doibase 10.1038/ncomms1990} {\bibfield
   {journal} {\bibinfo  {journal} {Nature Communications}\ }\textbf {\bibinfo
  {volume} {3}},\ \bibinfo {pages} {988} (\bibinfo {year} {2012})}\BibitemShut
  {NoStop}%
\bibitem [{\citenamefont {Yamada}\ \emph {et~al.}(2007)\citenamefont {Yamada},
  \citenamefont {Kasai}, \citenamefont {Nakatani}, \citenamefont {Kobayashi},
  \citenamefont {Kohno}, \citenamefont {Thiaville},\ and\ \citenamefont
  {Ono}}]{Yamada2007}%
  \BibitemOpen
  \bibfield  {author} {\bibinfo {author} {\bibfnamefont {K.}~\bibnamefont
  {Yamada}}, \bibinfo {author} {\bibfnamefont {S.}~\bibnamefont {Kasai}},
  \bibinfo {author} {\bibfnamefont {Y.}~\bibnamefont {Nakatani}}, \bibinfo
  {author} {\bibfnamefont {K.}~\bibnamefont {Kobayashi}}, \bibinfo {author}
  {\bibfnamefont {H.}~\bibnamefont {Kohno}}, \bibinfo {author} {\bibfnamefont
  {A.}~\bibnamefont {Thiaville}}, \ and\ \bibinfo {author} {\bibfnamefont
  {T.}~\bibnamefont {Ono}},\ }\href {\doibase 10.1038/nmat1867} {\bibfield
  {journal} {\bibinfo  {journal} {Nature Materials}\ }\textbf {\bibinfo
  {volume} {6}},\ \bibinfo {pages} {270} (\bibinfo {year} {2007})}\BibitemShut
  {NoStop}%
\bibitem [{\citenamefont {Kim}\ \emph {et~al.}(2007)\citenamefont {Kim},
  \citenamefont {Choi}, \citenamefont {Lee}, \citenamefont {Guslienko},\ and\
  \citenamefont {Jeong}}]{Kim2007}%
  \BibitemOpen
  \bibfield  {author} {\bibinfo {author} {\bibfnamefont {S.-K.}\ \bibnamefont
  {Kim}}, \bibinfo {author} {\bibfnamefont {Y.-S.}\ \bibnamefont {Choi}},
  \bibinfo {author} {\bibfnamefont {K.-S.}\ \bibnamefont {Lee}}, \bibinfo
  {author} {\bibfnamefont {K.~Y.}\ \bibnamefont {Guslienko}}, \ and\ \bibinfo
  {author} {\bibfnamefont {D.-E.}\ \bibnamefont {Jeong}},\ }\href {\doibase
  10.1063/1.2773748} {\bibfield  {journal} {\bibinfo  {journal} {Appl. Phys.
  Lett.}\ }\textbf {\bibinfo {volume} {91}},\ \bibinfo {pages} {082506}
  (\bibinfo {year} {2007})}\BibitemShut {NoStop}%
\bibitem [{\citenamefont {R\"{o}\ss~ler}\ \emph {et~al.}(2011)\citenamefont
  {R\"{o}\ss~ler}, \citenamefont {Leonov},\ and\ \citenamefont
  {Bogdanov}}]{Rossler2011}%
  \BibitemOpen
  \bibfield  {author} {\bibinfo {author} {\bibfnamefont {U.~K.}\ \bibnamefont
  {R\"{o}\ss~ler}}, \bibinfo {author} {\bibfnamefont {A.~A.}\ \bibnamefont
  {Leonov}}, \ and\ \bibinfo {author} {\bibfnamefont {A.~N.}\ \bibnamefont
  {Bogdanov}},\ }\href {http://stacks.iop.org/1742-6596/303/i=1/a=012105}
  {\bibfield  {journal} {\bibinfo  {journal} {J. Phys.: Conference Series}\
  }\textbf {\bibinfo {volume} {303}},\ \bibinfo {pages} {012105} (\bibinfo
  {year} {2011})}\BibitemShut {NoStop}%
\bibitem [{\citenamefont {Johnson}\ \emph {et~al.}(1996)\citenamefont
  {Johnson}, \citenamefont {Bloemen}, \citenamefont {den Broeder},\ and\
  \citenamefont {de~Vries}}]{Johnson1996}%
  \BibitemOpen
  \bibfield  {author} {\bibinfo {author} {\bibfnamefont {M.~T.}\ \bibnamefont
  {Johnson}}, \bibinfo {author} {\bibfnamefont {P.~J.~H.}\ \bibnamefont
  {Bloemen}}, \bibinfo {author} {\bibfnamefont {F.~J.~A.}\ \bibnamefont {den
  Broeder}}, \ and\ \bibinfo {author} {\bibfnamefont {J.~J.}\ \bibnamefont
  {de~Vries}},\ }\href {http://stacks.iop.org/0034-4885/59/i=11/a=002}
  {\bibfield  {journal} {\bibinfo  {journal} {Rep. Prog. Phys.}\ }\textbf
  {\bibinfo {volume} {59}},\ \bibinfo {pages} {1409} (\bibinfo {year}
  {1996})}\BibitemShut {NoStop}%
\bibitem [{\citenamefont {Yi}\ \emph {et~al.}(2009)\citenamefont {Yi},
  \citenamefont {Onoda}, \citenamefont {Nagaosa},\ and\ \citenamefont
  {Han}}]{Yi09}%
  \BibitemOpen
  \bibfield  {author} {\bibinfo {author} {\bibfnamefont {S.~D.}\ \bibnamefont
  {Yi}}, \bibinfo {author} {\bibfnamefont {S.}~\bibnamefont {Onoda}}, \bibinfo
  {author} {\bibfnamefont {N.}~\bibnamefont {Nagaosa}}, \ and\ \bibinfo
  {author} {\bibfnamefont {J.~H.}\ \bibnamefont {Han}},\ }\href {\doibase
  10.1103/PhysRevB.80.054416} {\bibfield  {journal} {\bibinfo  {journal} {Phys.
  Rev. B}\ }\textbf {\bibinfo {volume} {80}},\ \bibinfo {pages} {054416}
  (\bibinfo {year} {2009})}\BibitemShut {NoStop}%
\bibitem [{\citenamefont {Butenko}\ \emph {et~al.}(2010)\citenamefont
  {Butenko}, \citenamefont {Leonov}, \citenamefont {R\"o\ss{}ler},\ and\
  \citenamefont {Bogdanov}}]{Butenko2010}%
  \BibitemOpen
  \bibfield  {author} {\bibinfo {author} {\bibfnamefont {A.~B.}\ \bibnamefont
  {Butenko}}, \bibinfo {author} {\bibfnamefont {A.~A.}\ \bibnamefont {Leonov}},
  \bibinfo {author} {\bibfnamefont {U.~K.}\ \bibnamefont {R\"o\ss{}ler}}, \
  and\ \bibinfo {author} {\bibfnamefont {A.~N.}\ \bibnamefont {Bogdanov}},\
  }\href {\doibase 10.1103/PhysRevB.82.052403} {\bibfield  {journal} {\bibinfo
  {journal} {Phys. Rev. B}\ }\textbf {\bibinfo {volume} {82}},\ \bibinfo
  {pages} {052403} (\bibinfo {year} {2010})}\BibitemShut {NoStop}%
\bibitem [{\citenamefont {Wilson}\ \emph {et~al.}(2012)\citenamefont {Wilson},
  \citenamefont {Karhu}, \citenamefont {Quigley}, \citenamefont {R\"o\ss{}ler},
  \citenamefont {Butenko}, \citenamefont {Bogdanov}, \citenamefont
  {Robertson},\ and\ \citenamefont {Monchesky}}]{Wilson2012}%
  \BibitemOpen
  \bibfield  {author} {\bibinfo {author} {\bibfnamefont {M.~N.}\ \bibnamefont
  {Wilson}}, \bibinfo {author} {\bibfnamefont {E.~A.}\ \bibnamefont {Karhu}},
  \bibinfo {author} {\bibfnamefont {A.~S.}\ \bibnamefont {Quigley}}, \bibinfo
  {author} {\bibfnamefont {U.~K.}\ \bibnamefont {R\"o\ss{}ler}}, \bibinfo
  {author} {\bibfnamefont {A.~B.}\ \bibnamefont {Butenko}}, \bibinfo {author}
  {\bibfnamefont {A.~N.}\ \bibnamefont {Bogdanov}}, \bibinfo {author}
  {\bibfnamefont {M.~D.}\ \bibnamefont {Robertson}}, \ and\ \bibinfo {author}
  {\bibfnamefont {T.~L.}\ \bibnamefont {Monchesky}},\ }\href {\doibase
  10.1103/PhysRevB.86.144420} {\bibfield  {journal} {\bibinfo  {journal} {Phys.
  Rev. B}\ }\textbf {\bibinfo {volume} {86}},\ \bibinfo {pages} {144420}
  (\bibinfo {year} {2012})}\BibitemShut {NoStop}%
\bibitem [{\citenamefont {Huang}\ and\ \citenamefont
  {Chien}(2012)}]{Huang2012}%
  \BibitemOpen
  \bibfield  {author} {\bibinfo {author} {\bibfnamefont {S.~X.}\ \bibnamefont
  {Huang}}\ and\ \bibinfo {author} {\bibfnamefont {C.~L.}\ \bibnamefont
  {Chien}},\ }\href {\doibase 10.1103/PhysRevLett.108.267201} {\bibfield
  {journal} {\bibinfo  {journal} {Phys. Rev. Lett.}\ }\textbf {\bibinfo
  {volume} {108}},\ \bibinfo {pages} {267201} (\bibinfo {year}
  {2012})}\BibitemShut {NoStop}%
\bibitem [{\citenamefont {Tchoe}\ and\ \citenamefont {Han}(2012)}]{Tchoe12}%
  \BibitemOpen
  \bibfield  {author} {\bibinfo {author} {\bibfnamefont {Y.}~\bibnamefont
  {Tchoe}}\ and\ \bibinfo {author} {\bibfnamefont {J.~H.}\ \bibnamefont
  {Han}},\ }\href {\doibase 10.1103/PhysRevB.85.174416} {\bibfield  {journal}
  {\bibinfo  {journal} {Phys. Rev. B}\ }\textbf {\bibinfo {volume} {85}},\
  \bibinfo {pages} {174416} (\bibinfo {year} {2012})}\BibitemShut {NoStop}%
\bibitem [{\citenamefont {Bogdanov}\ and\ \citenamefont
  {Yablonskii}(1989)}]{Bogdanov89}%
  \BibitemOpen
  \bibfield  {author} {\bibinfo {author} {\bibfnamefont {A.~N.}\ \bibnamefont
  {Bogdanov}}\ and\ \bibinfo {author} {\bibfnamefont {D.~A.}\ \bibnamefont
  {Yablonskii}},\ }\href@noop {} {\bibfield  {journal} {\bibinfo  {journal}
  {Sov. Phys. JETP}\ }\textbf {\bibinfo {volume} {68}},\ \bibinfo {pages} {101}
  (\bibinfo {year} {1989})}\BibitemShut {NoStop}%
\bibitem [{\citenamefont {Bogdanov}\ and\ \citenamefont
  {Hubert}(1994)}]{Bogdanov94}%
  \BibitemOpen
  \bibfield  {author} {\bibinfo {author} {\bibfnamefont {A.}~\bibnamefont
  {Bogdanov}}\ and\ \bibinfo {author} {\bibfnamefont {A.}~\bibnamefont
  {Hubert}},\ }\href@noop {} {\bibfield  {journal} {\bibinfo  {journal} {J.
  Magn. Magn. Mater.}\ }\textbf {\bibinfo {volume} {138}},\ \bibinfo {pages}
  {255} (\bibinfo {year} {1994})}\BibitemShut {NoStop}%
\bibitem [{\citenamefont {R\"{o}\ss~ler}\ \emph {et~al.}(2006)\citenamefont
  {R\"{o}\ss~ler}, \citenamefont {Bogdanov},\ and\ \citenamefont
  {Pfleiderer}}]{Rosler2006}%
  \BibitemOpen
  \bibfield  {author} {\bibinfo {author} {\bibfnamefont {U.~K.}\ \bibnamefont
  {R\"{o}\ss~ler}}, \bibinfo {author} {\bibfnamefont {A.~N.}\ \bibnamefont
  {Bogdanov}}, \ and\ \bibinfo {author} {\bibfnamefont {C.}~\bibnamefont
  {Pfleiderer}},\ }\href {\doibase 10.1038/nature05056} {\bibfield  {journal}
  {\bibinfo  {journal} {Nature}\ }\textbf {\bibinfo {volume} {442}},\ \bibinfo
  {pages} {797} (\bibinfo {year} {2006})}\BibitemShut {NoStop}%
\bibitem [{\citenamefont {Han}\ \emph {et~al.}(2010)\citenamefont {Han},
  \citenamefont {Zang}, \citenamefont {Yang}, \citenamefont {Park},\ and\
  \citenamefont {Nagaosa}}]{Han10}%
  \BibitemOpen
  \bibfield  {author} {\bibinfo {author} {\bibfnamefont {J.~H.}\ \bibnamefont
  {Han}}, \bibinfo {author} {\bibfnamefont {J.}~\bibnamefont {Zang}}, \bibinfo
  {author} {\bibfnamefont {Z.}~\bibnamefont {Yang}}, \bibinfo {author}
  {\bibfnamefont {J.-H.}\ \bibnamefont {Park}}, \ and\ \bibinfo {author}
  {\bibfnamefont {N.}~\bibnamefont {Nagaosa}},\ }\href {\doibase
  10.1103/PhysRevB.82.094429} {\bibfield  {journal} {\bibinfo  {journal} {Phys.
  Rev. B}\ }\textbf {\bibinfo {volume} {82}},\ \bibinfo {pages} {094429}
  (\bibinfo {year} {2010})}\BibitemShut {NoStop}%
\bibitem [{\citenamefont {Dzyaloshinsky}(1958)}]{Dzyaloshinsky1958}%
  \BibitemOpen
  \bibfield  {author} {\bibinfo {author} {\bibfnamefont {I.}~\bibnamefont
  {Dzyaloshinsky}},\ }\href {\doibase 10.1016/0022-3697(58)90076-3} {\bibfield
  {journal} {\bibinfo  {journal} {J. Phys. Chem. Solids}\ }\textbf {\bibinfo
  {volume} {4}},\ \bibinfo {pages} {241} (\bibinfo {year} {1958})}\BibitemShut
  {NoStop}%
\bibitem [{\citenamefont {Moriya}(1960{\natexlab{a}})}]{Moriya60}%
  \BibitemOpen
  \bibfield  {author} {\bibinfo {author} {\bibfnamefont {T.}~\bibnamefont
  {Moriya}},\ }\href {\doibase 10.1103/PhysRev.120.91} {\bibfield  {journal}
  {\bibinfo  {journal} {Phys. Rev.}\ }\textbf {\bibinfo {volume} {120}},\
  \bibinfo {pages} {91} (\bibinfo {year} {1960}{\natexlab{a}})}\BibitemShut
  {NoStop}%
\bibitem [{\citenamefont {Moriya}(1960{\natexlab{b}})}]{Moriya60b}%
  \BibitemOpen
  \bibfield  {author} {\bibinfo {author} {\bibfnamefont {T.}~\bibnamefont
  {Moriya}},\ }\href {\doibase 10.1103/PhysRevLett.4.228} {\bibfield  {journal}
  {\bibinfo  {journal} {Phys. Rev. Lett.}\ }\textbf {\bibinfo {volume} {4}},\
  \bibinfo {pages} {228} (\bibinfo {year} {1960}{\natexlab{b}})}\BibitemShut
  {NoStop}%
\bibitem [{\citenamefont {Bazaliy}\ \emph {et~al.}(1998)\citenamefont
  {Bazaliy}, \citenamefont {Jones},\ and\ \citenamefont {Zhang}}]{Bazaliy98}%
  \BibitemOpen
  \bibfield  {author} {\bibinfo {author} {\bibfnamefont {Y.~B.}\ \bibnamefont
  {Bazaliy}}, \bibinfo {author} {\bibfnamefont {B.~A.}\ \bibnamefont {Jones}},
  \ and\ \bibinfo {author} {\bibfnamefont {S.-C.}\ \bibnamefont {Zhang}},\
  }\href {\doibase 10.1103/PhysRevB.57.R3213} {\bibfield  {journal} {\bibinfo
  {journal} {Phys. Rev. B}\ }\textbf {\bibinfo {volume} {57}},\ \bibinfo
  {pages} {R3213} (\bibinfo {year} {1998})}\BibitemShut {NoStop}%
\bibitem [{\citenamefont {Li}\ and\ \citenamefont {Zhang}(2004)}]{Li04}%
  \BibitemOpen
  \bibfield  {author} {\bibinfo {author} {\bibfnamefont {Z.}~\bibnamefont
  {Li}}\ and\ \bibinfo {author} {\bibfnamefont {S.}~\bibnamefont {Zhang}},\
  }\href {\doibase 10.1103/PhysRevLett.92.207203} {\bibfield  {journal}
  {\bibinfo  {journal} {Phys. Rev. Lett.}\ }\textbf {\bibinfo {volume} {92}},\
  \bibinfo {pages} {207203} (\bibinfo {year} {2004})}\BibitemShut {NoStop}%
\bibitem [{\citenamefont {Tatara}\ \emph {et~al.}(2008)\citenamefont {Tatara},
  \citenamefont {Kohno},\ and\ \citenamefont {Shibata}}]{Tatara2008}%
  \BibitemOpen
  \bibfield  {author} {\bibinfo {author} {\bibfnamefont {G.}~\bibnamefont
  {Tatara}}, \bibinfo {author} {\bibfnamefont {H.}~\bibnamefont {Kohno}}, \
  and\ \bibinfo {author} {\bibfnamefont {J.}~\bibnamefont {Shibata}},\ }\href
  {\doibase 10.1016/j.physrep.2008.07.003} {\bibfield  {journal} {\bibinfo
  {journal} {Phys. Rep.}\ }\textbf {\bibinfo {volume} {468}},\ \bibinfo {pages}
  {213} (\bibinfo {year} {2008})}\BibitemShut {NoStop}%
\bibitem [{not()}]{noteNumerics}%
  \BibitemOpen
  \href@noop {} {}\bibinfo {note} {We use dimensionless units in our
  simulations: length is in units of $J_{\rm{ex}}/D$; energy is in units of
  $J_{\rm{ex}}^2/D$; magnetic field is in units of $D^2/J_{\rm{ex}}$; time is
  in units of $J_{\rm{ex}}/(\gamma D^2)$; current is in units of $2D e/\hbar$;
  and voltage is in units of $\hbar\gamma D^2/(2eJ_{\rm{ex}})$. The system is
  discretized with grid size $0.2$; a smaller grid size is also used to check
  the accuracy of the results. The boundary condition is $\partial_{\mathbf{q}}
  \mathbf{n}=0$ where $\mathbf{q}$ is a vector normal to the boundary. To find
  the ground state, we anneal the system by adding a Gaussian noise field along
  the $z$ direction in $\mathbf{H}_{\rm{eff}}$. Equation \eqref{eq2} is solved
  by an explicit numerical scheme developed in Ref.
  \onlinecite{Serpico01}.}\BibitemShut {Stop}%
\bibitem [{\citenamefont {Altland}\ and\ \citenamefont
  {Simons}(2010)}]{SimonsQFT}%
  \BibitemOpen
  \bibfield  {author} {\bibinfo {author} {\bibfnamefont {A.}~\bibnamefont
  {Altland}}\ and\ \bibinfo {author} {\bibfnamefont {B.~D.}\ \bibnamefont
  {Simons}},\ }\href@noop {} {\emph {\bibinfo {title} {Condensed Matter Field
  Theory}}}\ (\bibinfo  {publisher} {Cambridge University Press},\ \bibinfo
  {address} {Cambridge},\ \bibinfo {year} {2010})\BibitemShut {NoStop}%
\bibitem [{\citenamefont {Lin}\ \emph {et~al.}(2013{\natexlab{b}})\citenamefont
  {Lin}, \citenamefont {Reichhardt}, \citenamefont {Batista},\ and\
  \citenamefont {Saxena}}]{szlin13skyrmion1}%
  \BibitemOpen
  \bibfield  {author} {\bibinfo {author} {\bibfnamefont {S.~Z.}\ \bibnamefont
  {Lin}}, \bibinfo {author} {\bibfnamefont {C.}~\bibnamefont {Reichhardt}},
  \bibinfo {author} {\bibfnamefont {C.~D.}\ \bibnamefont {Batista}}, \ and\
  \bibinfo {author} {\bibfnamefont {A.}~\bibnamefont {Saxena}},\ }\href@noop {}
  {\bibfield  {journal} {\bibinfo  {journal} {Phys. Rev. Lett.}\ }\textbf
  {\bibinfo {volume} {110}},\ \bibinfo {pages} {207202} (\bibinfo {year}
  {2013}{\natexlab{b}})}\BibitemShut {NoStop}%
\bibitem [{\citenamefont {Strogatz}(1994)}]{StrogatzBook}%
  \BibitemOpen
  \bibfield  {author} {\bibinfo {author} {\bibfnamefont {S.~H.}\ \bibnamefont
  {Strogatz}},\ }\href@noop {} {\emph {\bibinfo {title} {Nonlinear Dynamics And
  Chaos}}}\ (\bibinfo  {publisher} {Perseus Books},\ \bibinfo {address}
  {Reading, Massachusetts},\ \bibinfo {year} {1994})\BibitemShut {NoStop}%
\bibitem [{\citenamefont {Serpico}\ \emph {et~al.}(2001)\citenamefont
  {Serpico}, \citenamefont {Mayergoyz},\ and\ \citenamefont
  {Bertotti}}]{Serpico01}%
  \BibitemOpen
  \bibfield  {author} {\bibinfo {author} {\bibfnamefont {C.}~\bibnamefont
  {Serpico}}, \bibinfo {author} {\bibfnamefont {I.~D.}\ \bibnamefont
  {Mayergoyz}}, \ and\ \bibinfo {author} {\bibfnamefont {G.}~\bibnamefont
  {Bertotti}},\ }\href@noop {} {\bibfield  {journal} {\bibinfo  {journal} {J.
  Appl. Phys.}\ }\textbf {\bibinfo {volume} {89}},\ \bibinfo {pages} {6991}
  (\bibinfo {year} {2001})}\BibitemShut {NoStop}%
\end{thebibliography}
%

\end{document}